\documentstyle[aps,epsfig,graphicx]{revtex}
\begin{document}
\draft

\twocolumn[\hsize\textwidth\columnwidth\hsize\csname @twocolumnfalse\endcsname

\title{
Characterization of a Quasi-One-Dimensional Spin-1/2 Magnet
which is Gapless and Paramagnetic for $g\mu_BH\stackrel{<}{\textstyle 
\sim}J$ and $k_BT \ll J$
}
\author{P. R. Hammar$^1$, M. B. Stone$^1$, Daniel H. Reich$^1$,
C. Broholm$^{1,2}$, P. J. Gibson$^3$, M. M. Turnbull$^3$, C. P.
Landee$^3$, and M. Oshikawa$^{4}$}
\address{
$^1$Department of Physics and Astronomy, The Johns Hopkins
University, Baltimore, Maryland 21218\\
$^2$NIST Center for Neutron Research, National Institute of Standards
and Technology, Gaithersburg, Maryland
20899\\
$^3$Carlson School of Chemistry and Department of Physics,
Clark University, Worcester, Massachusetts 01610\\
$^{4}$Department of Physics, Tokyo Institute of Technology
Oh-oka-yama, Meguro-ku, Tokyo, 152-8551 JAPAN
}
\date{\today}
\maketitle
\begin{abstract}
High field magnetization, field-dependent specific heat measurements, 
and zero field inelastic magnetic neutron scattering have been used to 
explore the magnetic properties of copper pyrazine dinitrate ($\rm 
Cu(C_4H_4N_2)(NO_3)_2$).  The material is an ideal one-dimensional 
spin-1/2 Heisenberg antiferromagnet with nearest neighbor exchange 
constant $J=0.90(1)$ meV and chains extending along the orthorhombic 
$a$-direction.  As opposed to previously studied molecular-based 
spin-1/2 magnetic systems, copper pyrazine dinitrate remains gapless 
and paramagnetic for $g\mu_BH/J$ at least up to 1.4 and for $k_BT/J$ 
at least down to 0.03.  This makes the material an excellent model 
system for exploring the $T=0$ critical line which is expected in the 
$H-T$ phase diagram of the one-dimensional spin-1/2 Heisenberg 
antiferromagnet.  As a first example of such a study we present 
accurate measurements of the Sommerfeld constant of the spinon gas 
versus $g\mu_BH/J<1.4$ which reveal a decrease of the average spinon 
velocity by 32\% in that field range.  The results are in excellent 
agreement with numerical calculations based on the Bethe ansatz with 
no adjustable parameters.
\end{abstract}

\vskip2pc]

\narrowtext
\section{Introduction}
\label{sec:intro}

Quantum many body systems which support qualitatively different ground 
states as a function of Hamiltonian parameters have the potential to 
display so-called quantum critical phenomena when these parameters are 
tuned in the vicinity of zero temperature phase 
transitions.\cite{chakravarty,Sachdev97}  Many unusual physical 
phenomena such as heavy fermion behavior,\cite{Gabereview} high 
temperature superconductivity,\cite{Emery96,Castellani96} and 
non-fermi-liquid behavior\cite{Julian98} are now thought to be 
manifestations of proximity to such quantum critical points.  To 
understand the intrinsic properties of quantum criticality it is 
important to explore its phenomenology in simple and well controlled 
model systems.  One of the simplest quantum critical many body systems 
is a linear chain of antiferromagnetically coupled spin-1/2 objects.  
Quantum criticality in this system is particularly interesting because 
it is possible to continuously tune the  critical exponents 
by the application of a magnetic field.\cite{BIK,Fledderjohann97,Audet97}

\begin{figure}[h]
\centering
\includegraphics[scale=.55]{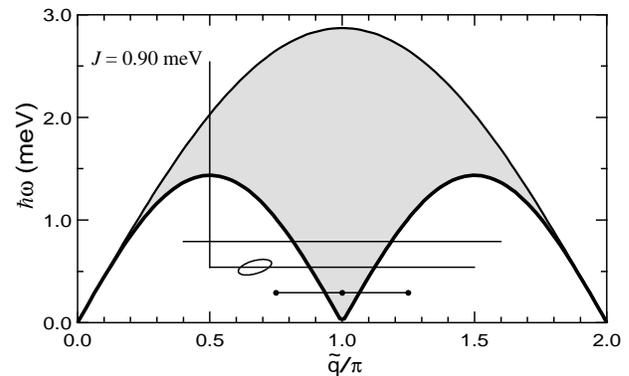}
\caption{Schematic view of the bounds of the spinon continuum of the
S=1/2 AFM chain, drawn for $J = 0.90$ meV as appropriate for
CuPzN. Solid dots show fixed values of $\tilde{q}$ for the data in
Fig.~\protect\ref{fig:Qperp}. Horizontal and vertical lines show
the scan directions for the data in Figs.~\protect\ref{fig:Qscans} and
\protect\ref{fig:Escan}.  The ellipse indicates the FWHM instrumental
resolution for Figs.~\protect\ref{fig:Qscans} and
\protect\ref{fig:Escan}.}
\label{fig:DispSketch}
\end{figure}

Unlike both conventional ordered magnets and gapped quantum spin 
chains which generally have sharp, low-energy dispersive modes, the 
excitation spectrum of the S=1/2 antiferromagnetic (AFM) chain is 
dominated by the so-called ``spinon continuum,'' 
\cite{Karbach97,Muller81} which is sketched in 
Fig.~\ref{fig:DispSketch}.  Some properties of the S=1/2 AFM chain may 
be computed exactly using the Bethe ansatz, \cite{betheansatz} but 
it is also very illuminating to map the spin chain onto a 1D 
system of interacting fermions.\cite{TsvelikBook}  This gives 
important insight into the spinon continuum, which may be viewed as 
the particle-hole continuum of the fermion model.  This underlying 
fermion character of the S=1/2 chain also has consequences for 
thermodynamic properties such as the magnetization and specific heat.  
There are a growing number of magnetic materials in which the 
predictions of these models have been tested 
experimentally.\cite{Heilmann78,Tennant93,Coldea97,Dender96,Dender97}

In the presence of a magnetic field $H$, an isotropic spin chain is 
described by the Hamiltonian
\begin{equation}
{\cal H} = \sum_i \left [ J {\bf S}_i \cdot {\bf S}_{i+1}
-g\mu_B {\bf H} \cdot {\bf S}_i \right ]  .
\end{equation}
The field causes a Zeeman splitting of the half-filled, 
doubly-degenerate spinon band, which introduces additional low-energy 
spanning vectors, and the spin chain has been predicted 
\cite{Muller81} to develop new soft modes at wave-vectors which are 
incommensurate with the lattice, but which connect the field-dependent 
Fermi points.  The dynamic critical exponents which control the 
temperature dependence of spin fluctuations close to these soft points 
are predicted to vary continuously with applied field.  In a recent 
experiment on copper benzoate we have obtained the first direct 
experimental evidence for the existence of these low-energy 
modes.\cite{Dender97}  However, while this work showed directly the 
field-induced incommensurability of the S=1/2 chain, it also showed a 
field-induced gap in copper benzoate.  Oshikawa and Affleck 
showed\cite{Oshikawa97} that this gap arises from the presence of two 
Cu sites per unit cell along the copper benzoate chains.  The 
combination of a staggered g-tensor and Dzyaloshinskii-Moriya 
interactions induces an effective staggered field when a uniform field 
is applied and this in turn creates a gap in the excitation spectrum 
and other interesting changes in the dynamic spin 
correlations.\cite{Oshikawa97,Essler98}

The result of this is that copper benzoate does not have a quantum 
critical line in the $H-T$ plane because a field immediately drives 
the system away from criticality.  In this paper we describe 
measurements on a long neglected one-dimensional spin-1/2 AFM, copper 
pyrazine dinitrate (CuPzN), which remains gapless for 
$g\mu_BH\stackrel{<}{\textstyle \sim}J$
and $k_BT \ll J$, and thus is very well suited for experimental studies 
of quantum criticality.  CuPzN is a S=1/2 chain with intra-chain 
exchange constant $J \approx 0.91$ meV ($J/k_{B} \approx 10.6$ K), 
as determined 
from previous zero-field susceptibility and specific 
heat studies.\cite{Losee73,Mennenga84}
It is highly one-dimensional; 3D ordering 
having not been observed in zero field for 
$T > 0.1$ K,\cite{Mennenga84} which implies \cite{Villain77} 
that the ratio of inter-chain to 
intra-chain coupling is $J'/J < 10^{-4}$.  The 
magnitude of the exchange constant $J$ in CuPzN makes the material an 
attractive candidate for both thermodynamic and neutron scattering 
studies in magnetic fields, as large effective fields $g \mu_{B}H/J$ 
can be obtained.  We report high field magnetization, field-dependent 
specific heat, and zero-field inelastic neutron scattering 
measurements that show that CuPzN is a nearly ideal example of an 
isotropic Heisenberg AFM chain.  In addition, our high field specific 
heat data provide direct experimental evidence for the suppression 
with field of the spinon velocity in the one-dimensional spin-1/2 AFM.

\section{Copper Pyrazine Dinitrate}

\begin{figure}[t]
\centering
\includegraphics[scale=.5]{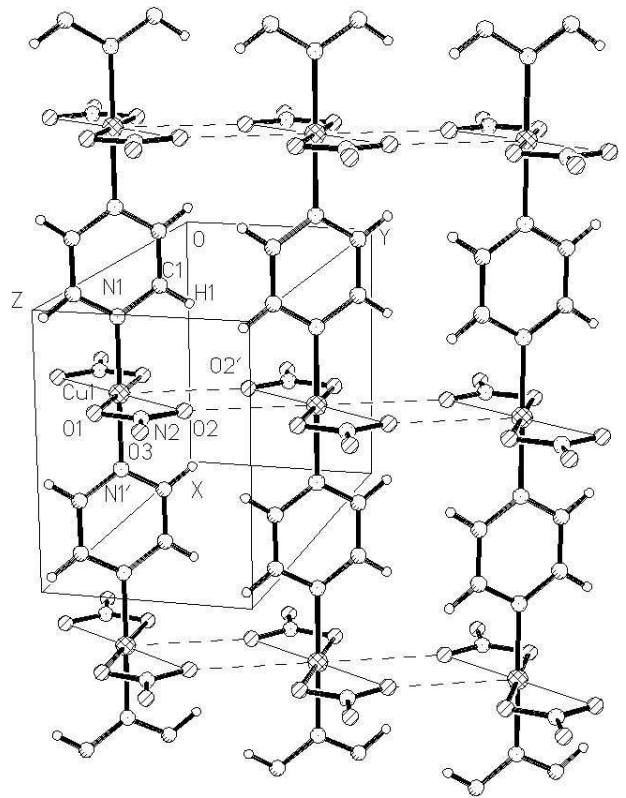}
\caption{Crystal structure of copper pyrazine dinitrate,
$\rm Cu(C_{4}H_{4}N_{2})(NO_{3})_{2}$, showing
how Cu$^{2+}$ ions (hatched spheres) are linked through pyrazine
rings to form 1D chains.
The chain axis ($\bf a$)
is vertical on the page, with the $\bf b$ axis nearly horizontal.
The long bonds
between the O2 atoms and the copper sites are represented by thin
lines
(intrachain contacts) and dashed lines (interchain contacts).
Atomic positions are given in Table 1, and
bond lengths and angles are given in Table 2.}
\label{fig:struct}
\end{figure}

CuPzN, $\rm Cu(C_{4}H_{4}N_{2})(NO_{3})_{2}$, is orthorhombic, with 
space group Pmna and room temperature lattice constants $a = 6.712$ 
\AA, $b = 5.142$ \AA, and $c = 11.732$ \AA.\cite{Santoro70} The 
copper ions form chains along the [100] direction, with one copper per 
unit cell along the chain.  The Cu ions along the chain are coupled 
magnetically through the pyrazine molecules, as shown in 
Fig.~\ref{fig:struct}.  The room temperature g-factors are $g_{a} = 
2.05$, $g_{b} = 2.27$, and $g_{c} = 2.07$.\cite{McGregor76}

We have recently re-determined the crystal structure 
of CuPzN with X-rays at $T = 158$ K.\cite{newstruct} 
 There are no major differences 
from the room temperature structure as originally determined.  
\cite{Santoro70} The cell parameters show very slight contraction upon 
cooling along the chain axis $a$ (6.712 \AA\ to 6.697 \AA\ ), and 
somewhat more change along the two axes perpendicular to the chains 
($b$ axis: 5.142 \AA\ to 5.112 \AA; $c$ axis: 11.732 \AA\ to 11.624 
\AA).  With neutrons, we have determined the low-temperature ($T < 4$ 
K) lattice constants to be $a = 6.6870$ \AA, and $b = 5.0983$ \AA. 
Throughout this paper we index wave vector transfer in the 
corresponding reciprocal lattice: ${\bf Q}=h{\bf a^*}+k{\bf b^*}+l{\bf 
c^*}$, where ${\bf a^*}=(2\pi/a)\hat{\bf a}$ etc.  We also define wave 
vector transfer along the chain as $\tilde{q}\equiv{\bf Q}\cdot{\bf 
a}=2\pi h$.

The atomic coordinates and some
 of the important bond lengths and bond angles measured at $T = 
158$ K are listed in Tables~\ref{tab:positions}
and \ref{tab:bonds}.  Each copper ion is 
coordinated to two nitrogen atoms from the adjacent pyrazine rings 
(Cu-N1 = 1.973 \AA) and two oxygen atoms (O1) from two different nitrate 
ions.  The Cu-O1 distance equals 2.004 \AA\, and is represented by a 
thick solid line in Fig.~\ref{fig:struct}.  The copper atoms and 
nitrate groups lie in mirror planes perpendicular to the two-fold $a$ axis. 
  There is a semi-coordinate bond between each 
copper atom and a second oxygen (O2) from each nitrate group, 
represented by a thin line in Fig.~\ref{fig:struct}.  The Cu-O2 
distance is 2.478 \AA, which is much greater than the 1.9 \AA\ for copper 
oxides with octahedral coordination, and the angle between the Cu-O2 
bond and the normal to the Cu1-N1-O1 plane 
is 30.4$^{\circ}$, which is 
also unfavorable for superexchange interactions.  The O2 atoms also 
bridge the copper atoms to the copper sites on adjacent chains by a 
long pathway (3.264 \AA, dashed lines). 
The axis connecting the Cu site to the O2$^{\prime}$ site on the adjacent
chain is tilted 
only 21.6$^{\circ}$ from the normal to the local coordination plane.  
This long interaction may help stabilize the formation of the three 
dimensional lattice, but there is no evidence that it provides a 
superexchange pathway.  The basal plane of the copper site is 
determined by the four short bonds.  The strength of the 
superexchange interactions through the pyrazine rings will be a 
function of the angle between the normal to the basal plane and the 
normal to the (planar) pyrazine molecules which is 51$^{\circ}$.  
\cite{Richardson76}

\begin{table}
{
\begin{tabular}{|c|c|c|c|c|}
Atom & x & y & z & U\\
\tableline
Cu1 & 0.0000 & 0.0000 & 0.0000 & 0.0071(3)\\
C1  & 0.3965(3) & -0.1709(5) & -0.0646(2) & 0.0097(4)\\
N1  & 0.2947(5) & 0.0000     & 0.0000     & 0.0086(7)\\
N2  & 0.0000    & -0.2599(6) & 0.1944(3)  & 0.0106(6)\\
O1  & 0.0000    & -0.0116(4) & 0.1723(3)  & 0.0104(6)\\
O2  & 0.0000    & -0.4137(6) & 0.1112(2)  & 0.0158(6)\\
O3  & 0.0000    & -0.3310(6) & 0.2952(2)  & 0.0208(6)\\
H1  & 0.319(5)  & -0.291(6)  & -0.105(3)  & 0.012\\
\end{tabular}
}
\caption{Fractional atomic coordinates and equivalent isotropic
thermal parameters for CuPzN at $T = 158$ K.}
\label{tab:positions}
\end{table}

\begin{table}
{
\begin{tabular}{|c|c|}
Bond Length & Bond Angle \\ \tableline
Cu1--N1 = 1.974(3) \AA &  N1--Cu1--O1 = 90$^{\circ}$\\
Cu1--O1 = 2.004(3) \AA &  O1--Cu1--O2 = 56.88(9)$^{\circ}$\\
Cu1--O2 = 2.478(3) \AA &  N1--Cu1--O2 = 90$^{\circ}$\\
\end{tabular}
}
\caption{Selected bond distances and angles for CuPzN at $T=158$ K}
\label{tab:bonds}
\end{table}

\section{Experimental Techniques}
\label{sec:exptech}

Crystals of CuPzN were grown by slow evaporation of aqueous solutions 
of Cu(II) nitrate and the corresponding pyrazine (either protonated or 
deuterated).  Large single crystals formed after several months.  
Faster evaporation results in the formation of many small, 
needle-shaped crystals.

Specific heat measurements on small, protonated crystals with typical 
mass $m = 5$ mg were made in fields up to $\mu_{0}H = 9$ Tesla for 0.1 
K $ < T <$ 10 K using the relaxation method.  
\cite{Bachmann72,hammarthesis} Magnetization measurements on both 
protonated and deuterated powder samples were carried out at the 
National High Field Magnet Laboratory in magnetic fields up to 
$\mu_{0}H = 30$ T using an EG\&G vibrating sample magnetometer which 
had been calibrated against a high-purity nickel sample.

For inelastic neutron scattering, a sample was used which consisted of 
two deuterated crystals with total mass $m = 72.8$ mg, co-aligned to 
within 0.4$^{\circ}$ in the $(hk0)$ scattering plane.  The 
measurements were performed on the SPINS cold neutron triple axis 
spectrometer at NIST. For measurements requiring a nearly isotropic wave 
vector resolution function we used a conventional parallel beam 
configuration.  In this configuration the projection of the 
instrumental resolution ellipsoid on the scattering plane is an 
ellipse with principal directions approximately parallel and 
perpendicular to the wave vector transfer $\bf Q$, and a Full Width 
at Half Maximum (FWHM) along these directions of typically $\Delta 
Q_{\parallel} = $ 0.02 \AA$^{-1}$ and $\Delta Q_{\perp} =$ 0.054 
\AA$^{-1}$ respectively.  For measurements requiring good $\bf Q$ 
resolution only along a single direction we used a horizontally 
focusing analyzer.  With this configuration the principal axes of the 
projection of the resolution ellipsoid on the scattering plane are 
approximately parallel and perpendicular to the direction of the 
scattered beam $\hat{\bf k}_f$, and the FWHM were $\Delta 
Q_{\parallel} = $ 0.045 \AA$^{-1}$, and $\Delta Q_{\perp} =$ 0.32 
\AA$^{-1}$ respectively.  By orienting the spin chain axis parallel to 
$\hat{\bf k}_f$ the instrument effectively integrates over wave vector 
transfer perpendicular to the chains.  This led to a gain in intensity 
by a factor of 5.3 over the conventional parallel beam configuration 
without significant loss in wave vector resolution along the chain 
axis.  The projection of the resolution ellipsoid on the energy 
transfer axis had FWHM $\Delta \hbar\omega = $ 0.15 meV in both 
configurations.  The half-value contour of the projection of the 
resolution function of the focusing analyzer configuration onto the 
$\hbar\omega$--$\tilde{q}$ plane is shown in 
Fig.~\ref{fig:DispSketch}.  We used 20 cm of beryllium oxide at $T=77$ 
K after the sample to reject neutrons with energies above the fixed 
final energy $E_f = 3.7$ meV from the detection system, and 10 cm of 
beryllium at 50 K before the sample to suppress higher order 
contamination of the incident beam for incident energies 
$E_i=E_f+\hbar\omega<5.1$ meV. The detector count rate was measured in 
units of the count rate of a monitor with sensitivity proportional to 
the incident wave length.  With this standard technique the raw 
count rates extracted from the experiment are proportional to the dynamic 
correlation function, ${\cal S}(Q,\omega )$.  For $E_i>5.1$ meV the 
beryllium filter prior to the sample was removed and corrections for 
higher order contributions to the monitor count rate were applied so 
as once again to yield a measurement of ${\cal S}(Q,\omega )$.

\section{Experimental Results}

\subsection{Magnetization}

\begin{figure}
\centering
\includegraphics[scale=.5]{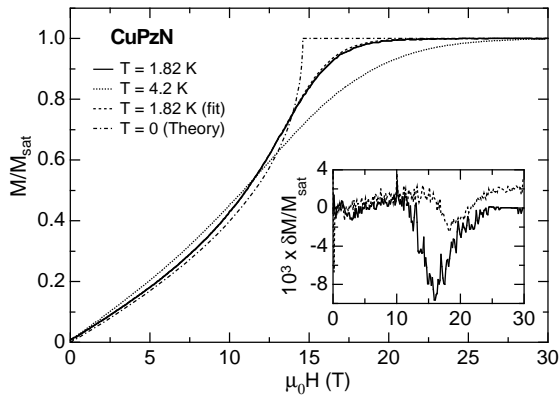}
\caption{Relative magnetization $M/M_{sat}$
of CuPzN at $T = 1.8$ K and 4.2 K.  Dot-dashed line:
calculated $T = 0$ magnetization (Ref.~\protect\cite{Griffiths64})
of an S=1/2 AFM chain with
$J/k_{B} = 10.3 $ K. Dashed line: fit of $T = 1.82 $ K data
to finite chain model described in text.  The fit of the $T = 4.2$ K
curve is indistinguishable from the data.  Inset: fit residuals
$\delta M = M - M_{fit}$.  Solid line: $T = 1.8$ K; dashed line:
$T = 4.2$ K.}
\label{fig:mag}
\end{figure}

The relative magnetization $M/M_{sat}$ of a protonated powder sample 
of CuPzN at $T = 4.2$ K and 1.8 K is shown in the main panel of 
Fig.~\ref{fig:mag} as a function of field up to $\mu_{0}H = 30 $ T. 
Results from a deuterated sample at $T = $ 4.2 K were 
indistinguishable from the protonated sample.  The data looks similar 
to previous magnetization data for other one-dimensional spin-1/2 
antiferromagnets.\cite{Mollymoto80}  The $T = 1.82$ K data set has an 
initial slope 14\% percent lower that of
the 4.2 K set, but demonstrates an increasingly positive curvature, 
crossing the higher temperature curve near 12 T and reaching 
saturation by 23 T. The data for both sets become measurably nonlinear 
with field for fields larger than 3 T. The trend of these data with 
decreasing temperature toward the indicated $T = 0$ theoretical curve 
\cite{Griffiths64} is clearly apparent.

\subsection{Specific Heat}

\begin{figure}[h]
\centering
\includegraphics[scale=.5]{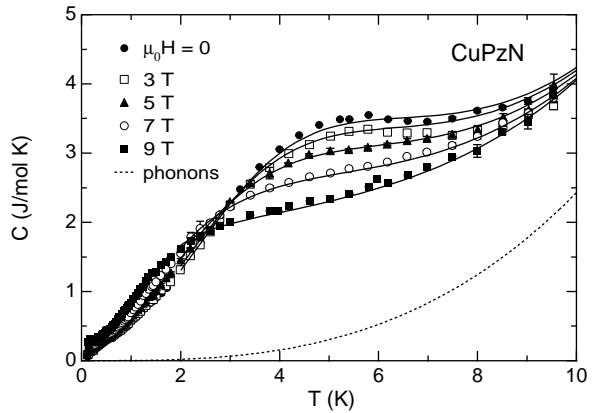}
\caption{Specific heat of CuPzN vs $T$ at constant magnetic field
$\mu_{0}H = 0$, 3 T, 5 T, 7 T and 9 T  $\parallel \hat{b}$.
Solid lines are a fit to an exact diagonalization
model as described in text. Dashed line is the phonon contribution
determined
from the fit.}
\label{fig:CvsH}
\end{figure}

Figure \ref{fig:CvsH} shows the temperature dependence of the total 
specific heat of CuPzN for a number of magnetic fields ${\bf H} 
\parallel {\bf b}$.  As $H$ is increased, the broad feature 
observed in zero-field is suppressed, and the maximum gradually shifts to 
lower $T$.  The solid lines are the results of a fit to a model based 
on exact diagonalization of short chains, as described below.  The 
dashed line is the estimated phonon contribution determined from the 
fit.

\begin{figure}[h]
\centering
\includegraphics[scale=0.5]{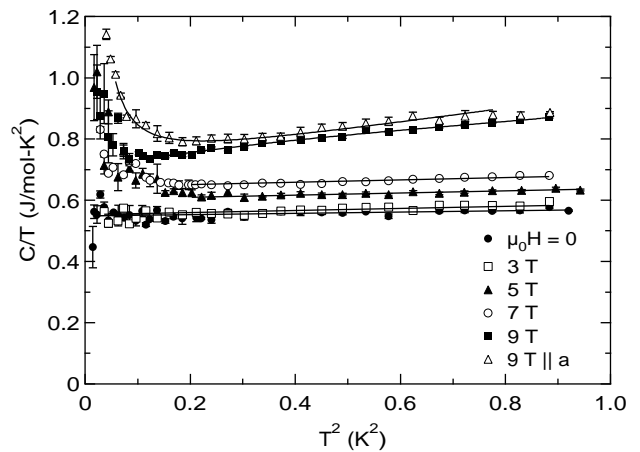}
\caption{Specific heat of CuPzN below $T = 1$ K,
plotted as $C/T$ vs $T^{2}$ for
at constant magnetic field
$\mu_{0}H = 0$, 3 T, 5 T, 7 T and 9 T  $\parallel \hat{b}$, and
$\mu_{0}H = 9$
T $\parallel \hat{a}$.}
\label{fig:lowTCp}
\end{figure}

Figure \ref{fig:lowTCp} shows an expanded view of the data in 
Fig.~\ref{fig:CvsH} for $T < 1 $ K, plotted as $C/T $ vs.  $T^{2}$.  
In addition, the specific heat at $\mu_{0}H = 9 $ T $\parallel 
{\bf a}$ is shown.  A predominantly linear in $T$ behavior is 
observed.  The positive slope visible at $\mu_{0}H = 9$ T is a 
consequence of the field-induced shift of spectral weight to lower 
temperature.  An increase in the linear part of $C(T)$ with increasing 
field is also seen.  This indicates a re-normalization of the spinon 
velocity with field, as will be discussed below.

\subsection{Neutron Scattering}

\begin{figure}[h]
\centering
\includegraphics[scale=0.5]{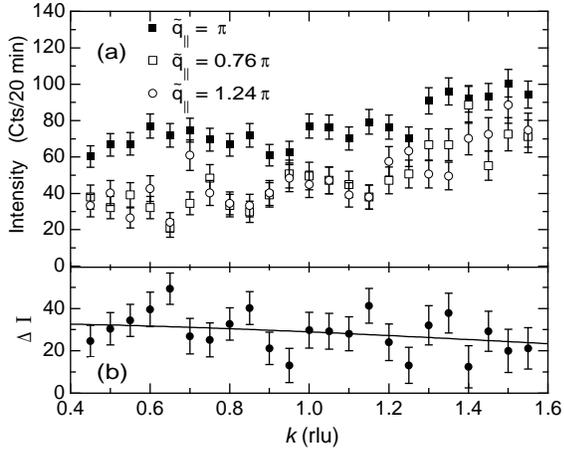}
\caption{
Neutron scattering intensity vs wave vector transfer $k\bf b^*$
perpendicular to
the chains at $\hbar\omega = 0.29$ meV  at $T = 0.3$ K.
(a) raw data at $h = 0.5$, 0.38, and 0.62 ($\tilde{q} =
\pi$, $0.76 \pi$, and $1.24 \pi$.)  (b) Difference $\Delta I$
between $h = 0.5$
scan and the average of the $h=0.38$ and $h=0.62$ scans.  The lack of
variation in $\Delta I$ with $k$ shows the one-dimensionality of the
magnetic scattering in CuPzN. }
\label{fig:Qperp}
\end{figure}

While bulk measurements have established that CuPzN is a 
quasi-one-dimensional magnet they cannot distinguish along which 
directions spins are coupled and along which directions they are not.  
For CuPzN both the $\bf a$ and $\bf b$ directions are candidate chain 
directions.  To distinguish between these possibilities we measured 
the wave vector dependence of the inelastic magnetic neutron 
scattering cross section in the (hk0) plane at low temperatures and 
for low values of energy transfer $\hbar\omega$.  
Fig.~\ref{fig:Qperp}(a) shows three scans along the $\bf b^*$ 
direction collected at $T = 0.3$ K and $\hbar\omega = 0.29$ meV, and 
for three values of wave vector transfer along the perpendicular $\bf 
a^*$ axis (indicated as dots in Fig.~\ref{fig:DispSketch}).  None of 
the scans shows any statistically significant modulation.  There is,
however, clearly more intensity for ${\bf Q }\cdot{\bf a^*}= \pi$ than 
for the other nearby values of this component of wave vector transfer.  
This is brought out by Fig.~\ref{fig:Qperp}(b) which shows the 
difference between the ${\bf Q }\cdot{\bf a^*}= \pi$ and the ${\bf Q 
}\cdot{\bf a^*}\ne \pi$ data.  No modulation beyond that associated 
with the single-ion squared magnetic form factor (solid line) is 
observed, which indicates the absence of spin correlations along the 
$\bf b^*$ direction on a time scale $1/\omega = 2$ ps for 
$k_BT\ll\hbar\omega $.  This observation allows us to use the 
horizontally focusing analyzer to gain sensitivity by probing the $\bf 
b^*$--integrated intensity as we explore the $\tilde{q}\equiv\bf 
Q\cdot a$ dependence of ${\cal S}({\bf Q} , \omega )$.

\begin{figure}[h]
\centering
\includegraphics[scale=0.5]{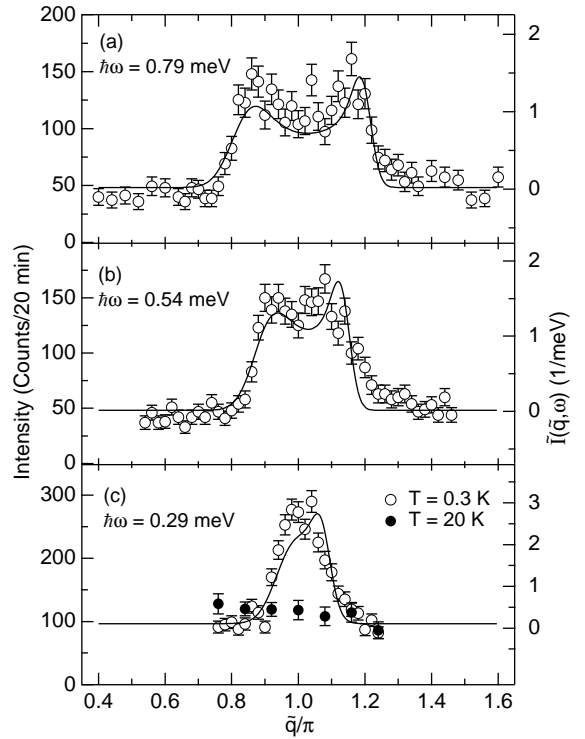}
\caption{
Neutron scattering intensity vs wave vector transfer along
the chains. Open symbols: $T = 0.3$ K.  Filled symbols: $T = 20$ K.
Right-hand axis gives normalized intensity scale
as described in text.  Solid lines are fit to a $T=0$ model
for ${\cal S}(\tilde{q},\omega)$ (Ref.~\protect\cite{Muller81}).}
\label{fig:Qscans}
\end{figure}

\begin{figure}[h]
\centering
\includegraphics[scale=0.5]{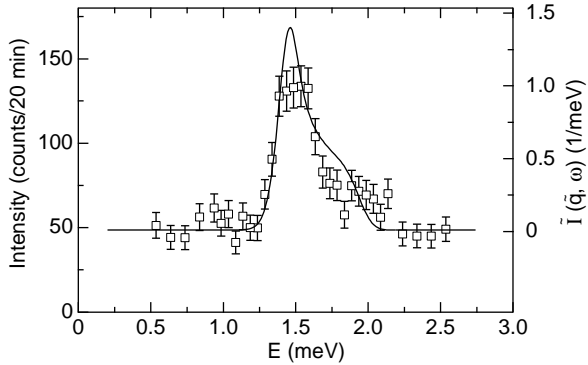}
\caption{Neutron scattering intensity vs energy transfer at
$\tilde{q} = \pi/2$ and $T = 0.3$ K. Right-hand axis gives
normalized intensity scale
as described in text. The solid line is fit to a $T=0$ model
for ${\cal S}(\tilde{q},\omega)$ (Ref.~\protect\cite{Muller81}).}
\label{fig:Escan}
\end{figure}

Data collected with the focusing analyzer are shown in 
Figs.~\ref{fig:Qscans} and \ref{fig:Escan}.  Figure \ref{fig:Qscans} 
shows the inelastic scattering intensity versus wave-vector transfer 
$\tilde{q}$ along the chains at $T = 0.3$ K for three fixed values of 
energy transfer $\hbar \omega = $ 0.29 meV, 0.54 meV, and 0.79 meV. 
Note that in all these scans wave vector transfer along the $\bf b^*$ 
direction was varied so as to maintain $\bf k_f \parallel a^*$ as 
required to maintain good wave vector resolution along the spin chain 
in the focusing analyzer configuration.  At $\hbar\omega = 0.29$ meV a 
single peak is observed at $\tilde{q} = \pi$, consistent with the data 
in Fig.~\ref{fig:Qperp}.  When the temperature is raised to $T = 20$ K 
this peak disappears, as shown by the solid symbols in 
Fig.~\ref{fig:Qscans}(c).  This indicates that the inelastic 
scattering observed is indeed magnetic in origin.  As $\hbar\omega$ is 
increased at $T = 0.3$ K two peaks become visible, demonstrating the 
presence of dispersion along the chain direction.  At $\hbar\omega = 
0.79$ meV, the two peaks are clearly resolved, but it is also apparent 
that the scattering cross section between the peaks is finite which 
indicates a continuum contribution.  The locations of the maxima in 
these constant-$\hbar\omega$ scans coincide with the  
lower bound of the spinon continuum anticipated for 
CuPzN on the basis of bulk measurements (see 
Fig.~\ref{fig:DispSketch}).  Figure~\ref{fig:Escan} shows the 
scattering intensity vs.  $\hbar\omega$ at $\tilde{q} = \pi/2$ (ie.  
along the vertical line in Fig.~\ref{fig:DispSketch}).  The abrupt 
increase in intensity at $\hbar\omega \approx 1.3$ meV and the 
subsequent gradual decay at higher energies are again characteristic 
of the bounded spinon continuum.  The solid lines in 
Figs.~\ref{fig:Qscans} and \ref{fig:Escan} were obtained from a fit to 
a model of the two-spinon contributions to ${\cal S}(\tilde{q},\omega 
)$ to be described below.

\section{Discussion}

\subsection{Magnetization}

As is well known, many of the finite temperature thermodynamic 
properties of quantum spin chains can be calculated quite accurately 
from exact diagonalization of short chains.\cite{Bonner_Fisher64} 
Comparison of data to calculations of this type is a very effective 
way of ascertaining how well a material is described by a particular 
model, and also of determining the exchange constants in that model.  
We now compare our results for the magnetization and specific heat of 
CuPzN to exact diagonalizations of linear chains containing up to $N = 
16$ spins-1/2.

For the magnetization, we find surprisingly good agreement between our 
data and the calculated result for $N = 16$.  We carried out a 
simultaneous fit to both data sets shown in Fig.~\ref{fig:mag} ($T = 
1.82$ and 4.2 K), using the measured average g-factor $g_{avg} = 
2.11$.  This fit gave $J/k_{B} = 10.3(1)$ K.\cite{magfitnote} The 
results of the fit are superimposed on the data in Fig.~\ref{fig:mag} 
but can only be distinguished from the data near $\mu_{0} H = 15$ T at $T = 
1.82$ K.  The residuals $\delta M = M - M_{fit}$ are shown in the 
inset of Fig.~\ref{fig:mag}.  These show the deviation of model from 
data to be significantly less than 1\% over most of the field range.  
As a measure of the error in this fit associated with the finite 
system size used in the calculation, we have studied the convergence 
of $M(H,T,N)$ with increasing $N$.  At $T = 1.82$ K ($k_{B}T/J = 
0.17$), the difference $[M(N = 16) - M(N=15)]/M_{sat} $ is less than 
0.003 for all $H$.  At $T = 4.2 $ K, this difference is less than $2 
\times 10^{-5}$.  We conclude that our exact diagonalization results 
are representative of infinite length chains in the parameter range 
studied and that the magnetization of CuPzN down to $k_BT/J=0.17$ is 
perfectly described by the linear chain model.

\subsection{Specific Heat}

At the high end of the temperature range accessed in our
measurements, the specific heat 
may also be calculated accurately from exact diagonalization.  
Following Bonner and Fisher,\cite{Bonner_Fisher64} we compute the 
specific heat at constant field $C_{H}(J,T)$ as the average of that 
obtained from chains with 15 and 16 spins.  Finite-size scaling 
suggests that this procedure gives quite accurate results down to 
temperatures $k_BT/J \approx 0.2 $.  All the data for $T > 2 $K in 
Fig.~\ref{fig:CvsH} were fit simultaneously to
\begin{equation}
C(T) = A C_{H}(J,T) + B T^{3} ,
\end{equation}
where the second term accounts for the lattice contribution.  As shown 
by the solid lines in Fig.~\ref{fig:CvsH}, this model accounts for the 
data very well, with $A = 1.05(1)$, $J/k_{B} = 10.6(1)$ K, and $B = 
2.4(1) $ mJ/mole-K$^{4}$.  We ascribe the deviation of $A$ from 1 to a 
systematic error in the normalization of our specific heat data.  This 
is corroborated by a computation of the magnetic entropy from the 
specific heat data, which comes out to be 4\% higher than the 
expected value of $R \ln 2$.\cite{hammarthesis}

\begin{figure}[h]
\centering
\includegraphics[scale=0.5]{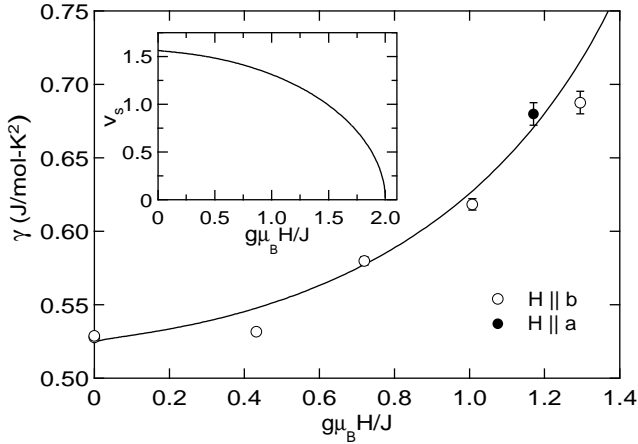}
\caption{Field dependence of the Sommerfeld constant
($\gamma=\lim_{T\rightarrow 0} C/T$)derived from  low-T  specific
heat data,
compared to Bethe ansatz
calculation (solid line).  The data was scaled by a factor 0.95 to
account for
a systematic error in normalization which was derived from the
comparison to
exact diagonalization data shown in Fig.~\protect\ref{fig:CvsH}.
Inset: calculated field-dependence of the spinon velocity $v_{s}(H)$.}
\label{fig:Cpgamma}
\end{figure}

The linear dispersion relation of the fermionic spinons at low energy 
implies that the low-temperature specific heat of the S=1/2 AFM chain 
should be linear in $T$.  It is  given by \cite{Blote86,Affleck86}
\begin{equation}
C_{H}(T) = \gamma_{H} T ~=~ \frac{\pi}{3}R \frac{k_{B}T}{v_{s}(H)} ,
\end{equation}
where $v_{s}(H)$ is the field-dependent spinon velocity.  In zero 
field, $v_{s} = \pi J/2$, and the specific heat becomes $C = (2/3)R 
k_{B}T/J$, \cite{Takahashi73,Johnson72} a result which has been 
confirmed experimentally in both copper 
benzoate \cite{Dender97} and CuPzN.\cite{Mennenga84}  The field-dependence of 
$v_{s}(H)$ can be determined from the Bethe 
ansatz \cite{BIK,KBIbook} and used to predict 
$C_{H}(T)$.  It is calculated as
\begin{equation}
	v_s  = \frac{E}{2 \pi \sigma(\Lambda)} ,
\end{equation}
where $E$ and $\sigma$ are determined by the following set of integral 
equations
\begin{eqnarray*}
\xi(\eta)
      &=& 1 - \frac{1}{2\pi}
         \int_{-\Lambda}^{\Lambda} K(\eta-\eta') \xi(\eta') d \eta' ,
				\\
\sigma(\eta)
      &=& \frac{1}{\pi ( 1 + \eta^2)}
	- \frac{1}{2\pi} \int_{-\Lambda}^{\Lambda}
		K(\eta - \eta') \sigma(\eta') d \eta'  , \\
\rho(\eta)
      &=& \frac{1}{2 \pi} \left[
                 \frac{d K(\eta - \Lambda)}{d \eta}
                 - \int_{-\Lambda}^{\Lambda} K(\eta - \eta')
			\rho(\eta') d \eta' \right],
				\nonumber\\
E     &=& \frac{4 \Lambda}{(1+\Lambda^2)^2}
          +\int_{-\Lambda}^{\Lambda}\epsilon_0(\eta)\rho(\eta) d\eta ,
\end{eqnarray*}
with $K(\eta) = 4/(4+\eta^2)$.
The boundary of integration $\Lambda$ is related to the magnetic
field $H$ by $g \mu_B H / J = 2 \pi \sigma(\Lambda)/\xi(\Lambda)$.
In practice, we first fixed the value of $\Lambda$, and then 
solved the integral equations by numerical integration and 
iteration  to obtain $v_s$ and $H$.  Accuracy of more than six digits 
can be achieved within minutes on a standard workstation.  We 
performed the calculation for many values of $\Lambda$ to give $v_s$ 
as a smooth function of $H$, which is
shown in the inset to Fig.~\ref{fig:Cpgamma}.
This determines the theoretical curve of 
the coefficient $\gamma(H)$ with no adjustable parameters, once
$J$ and the $g$-value are fixed.

Figure \ref{fig:Cpgamma} shows the measured field dependence of 
$\gamma_{H} = C_{H}/T$ determined from the data in 
Fig.~\ref{fig:lowTCp}, together with the calculated result.  These are 
plotted against reduced field $H^{*} = g \mu_{B} H/J$ with $J/k_{B} = 10.6$ 
K, and $g_{a} = 2.05$ and $g_{b} = 2.27$. \cite{McGregor76}  Note that 
for this comparison a correction factor $1/A=0.95$ was applied to the 
specific heat data to account for the normalization error discovered 
when comparing to exact diagonalization calculations.  The agreement 
between model and data is excellent and the data provides direct 
experimental evidence for the field-dependence of the spinon velocity 
in the S=1/2 AFM chain.

\subsection{Neutron Scattering}

The magnetic contribution to the measured inelastic neutron scattering 
signal is proportional to the normalized magnetic scattering intensity
\begin{equation}
\tilde{I}({\bf Q}, \omega )=|\frac{g}{2}F(Q)|^{2}~
\sum_{\alpha\beta}(\delta_{\alpha\beta}-
{\bf \hat{Q}_\alpha\hat{Q}_\beta })
\tilde{{\cal S}}^{\alpha\beta}({\bf Q}, \omega ).
\label{eq:I_tilde}
\end{equation}
In this expression $F(Q)$ is the magnetic form factor of Cu$^{2+}$, 
\cite{formfac} and $\tilde{{\cal S}}^{\alpha\beta}({\bf Q}, \omega )$ 
is the convolution of the dynamic spin correlation function, 
\cite{Lovesey}
\begin{equation}
{\cal S}^{\alpha\beta}({\bf Q}, \omega )=
\frac{1}{2\pi\hbar }
\int dt e^{i\omega t }\frac{1}{N}\sum_{\vec{R}\vec{R}^\prime}
<S_{\vec{R}}^\alpha (t) S_{\vec{R}^\prime}^\beta (0)>
e^{-i\vec{Q}\cdot
(\vec{R}-\vec{R}^\prime)} ,
\label{eq:Sqw}
\end{equation}
with the normalized instrumental resolution function. 
\cite{Chesser_Axe}  We determined $\tilde{I}({\bf Q}, \omega )$ in 
absolute units by comparison of the magnetic scattering intensity to 
the incoherent scattering intensity $I_{V}$ from a vanadium reference 
sample:
\begin{equation}
\tilde{I}({\bf Q}, \omega ) =
\left[ \frac{1}{\int d\hbar\omega I_{V}} \frac{n_{V}}{n_{Cu}}
(\frac{b_{i}}{r_{0}})^{2} \right ] I_{mag}({\bf Q}, \omega ) .
\label{eq:VanNorm}
\end{equation}
Here $n_{V}$ and $n_{Cu}$ are the number of moles of vanadium and 
copper in the reference and sample, respectively, $b_{i} = 6.36$ fm is 
the incoherent scattering length for vanadium, \cite{Lovesey} $r_{0} = 
5.38 $ fm, and $I_{mag}({\bf Q}, \omega )$ is the portion of the 
measured signal attributable to magnetic scattering.  The conversion 
of the measured count rates to $\tilde{I}({\bf Q}, \omega )$ is shown 
on the right-hand axis in Figs.~\ref{fig:Qscans} and \ref{fig:Escan}.  
As a complete, independent measure of the non-magnetic background was 
not made, the zero of the $\tilde{I}({\bf Q}, \omega )$ scale was set 
from the backgrounds determined in fitting the data.

M\"{u}ller {\em et al.} \cite{Muller81} proposed the following 
approximate expression for the $T=0$ dynamic spin correlation function 
for the spin-1/2 Heisenberg AFM chain,
\begin{equation}
{\cal S}^{\alpha\alpha}(\tilde{q},\omega) = \frac{1}{2\pi}\frac{\tilde{A}}
{\sqrt{(\hbar\omega)^2 - \epsilon_{1}^{2}(\tilde{q})}}
\Theta(\hbar\omega - \epsilon_{1}(\tilde{q}))
\Theta(  \epsilon_{2}(\tilde{q}) - \hbar\omega).
\label{eq:muller}
\end{equation}
In this expression $\Theta(x)$ is the step function, while 
$\epsilon_{1}(\tilde{q}) = (\pi J/2) |\sin{\tilde{q}}|$ and 
$\epsilon_{2}(\tilde{q}) = \pi J \sin{\tilde{q}/2} $ are the lower and 
upper bounds of the spinon continuum.  Recent publications present 
exact calculations of the contribution to ${\cal S}(\tilde{q},\omega 
)$ from n-spinon excitations. \cite{Karbach97,Bougourzi97,Abada97} 
The numerical differences between the exact expressions for ${\cal 
S}(\tilde{q},\omega)$ and the simpler approximate form are small, and 
are unimportant for the analysis of our relatively low statistics 
scattering data for CuPzN.  For the purpose of determining whether the 
scattering data can be described by the spin chain model which 
accounted for our bulk data we have therefore used Eq.  
\ref{eq:muller}.  This expression was convolved with the experimental 
resolution function and the overall scale factor, $\tilde{A}$, the 
exchange constant, $J$, and a constant background for each scan were 
varied to achieve the best possible fit to the scattering data.  The 
results of this fit with $\tilde{A} = 3.5(1)$ and $J = 0.90(1)$ meV are 
shown as solid lines in Figs.~\ref{fig:Qscans} and \ref{fig:Escan}, 
and may be seen to provide an excellent description of the data.  The 
value of $J$ derived from neutron scattering is indistinguishable from 
that derived from specific heat data.

\section{Conclusion}

The combination of magnetization, specific heat and inelastic neutron 
scattering reported here show that CuPzN is extremely well described 
by the simple isotropic Heisenberg model of the S=1/2 AFM spin chain 
with $J = 0.90(1)$ meV.  The neutron scattering measurements have 
confirmed that the crystalline $\bf a$-axis is the 1D chain axis, as 
had been predicted from structural analysis.  The field-dependence of 
the specific heat shows that CuPzN remains gapless in a field and that 
the spinon velocity is reduced in agreement with predictions based on 
calculations using the Bethe ansatz technique.  Taken as a whole, 
these results establish that CuPzN is an excellent model system for 
the one-dimensional spin-1/2 Heisenberg antiferromagnet even in large 
magnetic fields ($g\mu_BH/J\stackrel{<}{\textstyle \sim}1.4$) and at 
low temperatures ($k_BT/J\stackrel{>}{\textstyle \sim}0.03$).  This 
material is therefore an excellent candidate for further studies of 
the unique field dependent quantum critical properties of the spin-1/2 
Heisenberg antiferromagnet.

\section{Acknowledgments}

We thank Ian Affleck for stimulating discussions.
NSF grants DMR-9357518 and DMR-9453362 supported work at JHU.  This 
work utilized neutron research facilities supported by NIST and the 
NSF under Agreement No.  DMR-9423101.  A portion of this work was 
performed at the National High Magnetic Field Laboratory, which is 
supported by NSF cooperative agreement DMR-9527035 and the State of 
Florida.  DHR acknowledges support from the David and Lucile Packard 
Foundation.


\begin{references}
\bibitem{newstruct}
    Structural determination parameters: $a = 6.697(1)$ \AA, $ b = 
    5.112(1)$ \AA, $c = 11.624(3)$ \AA\ at T = 158 K, Z = 2, space group 
    Pmna.  A total of 888 reflections were observed, of which 619 were 
    unique.  R = 0.0508 based on the 541 reflections with $I >4\sigma(I)$.  
    R = 0.0564 based on all unique reflections.
\bibitem{magfitnote}
    This was actually a three parameter fit. In addition to $J$ it was 
    necessary to include offsets to account for small, field-independent 
    backgrounds arising from offset voltages in the measuring electronics.  
    These backgrounds are $0.007 M_{sat}$ and $0.006 M_{sat}$ for the $T= 
    1.82$ and 4.2 K data, respectively.
\end{references}
\end{document}